\documentclass[pra,showpacs,amsmath,amssymb, twocolumn, 10pt]{revtex4}

\usepackage{amsthm}
\usepackage{dcolumn}
\usepackage{bm}
\usepackage{graphicx}

\begin{document}

\newtheorem{corollary}{Corollary}
\newtheorem{definition}{Definition}
\newtheorem{example}{Example}
\newtheorem{lemma}{Lemma}
\newtheorem{proposition}{Proposition}
\newtheorem{theorem}{Theorem}
\newtheorem{fact}{Fact}
\newtheorem{property}{Property}
\newcommand{\bra}[1]{\langle #1|}
\newcommand{\ket}[1]{|#1\rangle}
\newcommand{\braket}[3]{\langle #1|#2|#3\rangle}
\newcommand{\ip}[2]{\langle #1|#2\rangle}
\newcommand{\op}[2]{|#1\rangle \langle #2|}

\newcommand{\tr}{{\rm tr}}
\newcommand{\supp}{{\it supp}}
\newcommand{\sch}{{\it Sch}}

\newcommand {\E } {{\mathcal{E}}}
\newcommand {\F } {{\mathcal{F}}}
\newcommand {\diag } {{\rm diag}}

\title{Local distinguishability of orthogonal $2\otimes3$ pure states}
\author{Yu Xin$^{1,2}$}
\email{xiny05@mails.tsinghua.edu.cn}
\author{Runyao Duan$^{1}$}
\email{dry@tsinghua.edu.cn}

\affiliation{$^1$State Key Laboratory of Intelligent Technology and
Systems, Tsinghua National Laboratory for Information Science and
Technology, Department of Computer Science and Technology, Tsinghua
University, Beijing 100084, China,
\\
$^2$Department of Physics, Tsinghua University, Beijing 100084,
China}

\date{\today}

\begin{abstract}
We present a complete characterization for the local
distinguishability of orthogonal $2\otimes 3$ pure states except for
some special cases of three states. Interestingly, we find there is
a large class of four or three states that are indistinguishable by
local projective measurements and classical communication (LPCC),
but can be perfectly distinguishable by LOCC. That indicates the
ability of LOCC for discriminating $2\otimes 3$ states is strictly
more powerful than that of LPCC, which is strikingly different from
the case of multi-qubit states. We also show that classical
communication plays a crucial role for local distinguishability by
constructing a class of $m\otimes n$ states which require at least
$2\min\{m,n\}-2$ rounds of classical communication in order to
achieve a perfect local discrimination.
\end{abstract}


\maketitle

\section{Introduction}
The basic problem for distinguishing quantum states by local
operations and classical communication (LOCC) can be formulated as
follows. Suppose two spatially separated parties, say Alice and Bob,
share a quantum state which is secretely chosen from a finite set of
pre-specified  quantum states. They want to figure out the identity
of the unknown state, but are only allowed to manipulate their own
quantum systems and to communicate with each other using classical
channels. This problem has received considerable attentions and has
been studied extensively. Numerous interesting results have been
reported, see Refs. \cite{PW91,BDF+99, BDM+99, DMS+00,WSHV00,
VSPM01, CY01,HM03, JCY05, ABB+05, OGA06, TDL01, DLT02, EW02, GKR+01,
FAN04, NAT05, OH06,GKR+02, GV01, WH02,CHE04,HSSH03,CL04,DER04,
WAT05, HMM+06, BW06, DFJY07,WS07, COH07a, DFXY07, FS07, COH07b,
OH07, DY07} for an incomplete list. Despite these exciting
progresses, it remains unknown how to determine the local
distinguishability of a set of multipartite states.

For the convenience of the readers, we give a brief review for some
of these results. Walgate and coworkers showed that any two
orthogonal pure states, no matter entangled or not, can always be
perfectly distinguishable by LOCC \cite{WSHV00}. Furthermore, it has
been shown that the local distinguishability and the global
distinguishablity of two pure states have the same efficiencies
\cite{VSPM01, CY01, HM03,JCY05, ABB+05, OGA06}. However, the
situation changes dramatically for a set of orthogonal states with
three or more members, where a perfect discrimination is generally
impossible. The most surprising discovery on this topic is that
there exists a set of nine $3\otimes 3$ orthogonal pure product
states which are indistinguishable by LOCC, a phenomenon known as
``nonlocality without entanglement" \cite{BDF+99, BDM+99, DMS+00}.
Inspired by this discovery, many researchers devoted to the local
distinguishability of product states. It is now clear that any set
of $2\otimes n$ orthogonal product pure states are perfectly
distinguishable by LOCC, but a set of incomplete orthogonal product
states which cannot be extended by adding some additional orthogonal
product state (UPB) is indistinguishable by LOCC \cite{BDM+99,
DMS+00}. The problem of distinguishing a complete basis has been
completely solved \cite{HSSH03,CL04, DER04}, but only very recently
a characterization for the locally distinguishable $3\otimes 3$
product states was obtained by Feng and Shi \cite{FS07}.

One of the main difficulties in studying local distinguishability is
that there is no effective characterization of LOCC operations. In
order to partially overcome this obstacle, many researchers began to
employ separable operations instead of LOCC operations to study the
local distinguishability. The effectiveness of this method can be
roughly  understood as follows. First, the class of separable
operations has a rather beautiful mathematical structure.  It is
much easier to  work with separable operations rather than LOCC
operations. Second, the class of LOCC operations is a subset of the
class of separable operations \cite{BDF+99}. So one can obtain
useful necessary conditions about local distinguishability  by
applying separable operations. Third, any separable operation can be
implemented by some LOCC operation with a nonzero success
probability. In other words, separable operations and LOCC
operations are probabilistically equivalent. Due to these reasons,
separable operations have been widely used in studying local
distinguishability. We shall briefly review two kinds of results:
probabilistic discrimination and perfect discrimination. Chefles
first studied the distinguishability of a set of general quantum
states by probabilistic LOCC and presented a necessary and
sufficient condition for the unambiguous distinguishability
\cite{CHE04}. A simplified version of this condition when only pure
states are under consideration was independently obtained by
Bandyopadhyay and Walgate \cite{BW06}, with which it was
demonstrated that any three pure states are distinguishable by
stochastic LOCC. Based on these results, Duan {\it et al} studied
the local distinguishability of an arbitrary basis of a multipartite
state space and provided a universal tight lower bound on the number
of locally unambiguously distinguishable members in an arbitrary
basis \cite{DFJY07}. Walgate and Scott further showed that this
lower bound plays a crucial role in deciding the generic properties
such as local unambiguous distinguishability of a set of randomly
chosen states \cite{WS07}. Separable operations were also used to
show that certain set of states are not perfectly distinguishable by
LOCC. More precisely, Nathanson showed that any $(d+1)$ maximally
entangled bipartite states on $d\otimes d$ cannot be perfectly
distinguishable by separable operations, thus also are
indistinguishable by LOCC \cite{NAT05}. The same result was
independently obtained  by Owari and Hayashi using a slightly
different method \cite{OH06}. It is interesting that before this
result Ghosh {\it et al} and Fan have respectively solved the
special cases of $d=2$ and $d=3$ using a rather different approach
\cite{GKR+01, FAN04}. Watrous constructed a class of bipartite
subspaces having no basis distinguishable by separable operations,
thus solved an open problem concerning with the environment-assisted
capacity of quantum channels \cite{WAT05}. Hayashi {\it et al}
studied the relation between average entanglement degree and local
distinguishability of a set of orthogonal states, and provided a
very general bound on the number of states which can be locally
distinguishable. In Ref. \cite{DFXY07} we systematically studied the
distinguishability of quantum states by separable operations and
found a new characterization for the distinguishability of quantum
states by separable operations. Notably, we showed that separable
operations acting on two-qubit are  strictly powerful than LOCC
operations. A more general class of locally indistinguishable
subspaces was also constructed.

All of the above works suggest that the problem of deciding the
local distinguishability of a set of general quantum states is
rather complicated. Interestingly, for some very simple cases such
as two-qubit states, an analytical solution is possible. Ghosh {\it
et~al} first obtained some partial results on the local
distinguishability of two-qubit states using some bounds on
entanglement distillation \cite{GKR+02}. Based on an idea of
Groisman and Vaidman \cite{GV01}, Walgate and Hardy obtained a very
simple characterization for a set of $2\otimes n$ orthogonal pure
states to be perfectly distinguishable by LOCC if the owner of the
qubit makes the first nontrivial measurement \cite{WH02}. Employing
this condition, they finally settled the local distinguishability of
$2\otimes2$ states \cite{WH02}. Another immediate consequence is
that local projective measurements and classical communication
(LPCC) is sufficient for the local distinguishability of multi-qubit
states \cite{DY07}, which greatly simplifies the local
distinguishability of multi-qubit states. But this is not true in
general. In Ref. \cite{BDM+99} Bennett and coworkers constructed a
set of five $3\otimes 4$  pure product states which are perfectly
distinguishable by LOCC but not by LPCC. Very recently a set of
$3\otimes 3$ states with similar property was obtained by Cohen
\cite{COH07a}. However, we still don't know whether the general POVM
is required in order to distinguish $2\otimes 3$ states. It seems
somewhat strange that the local distinguishability of $2\otimes 3$
states when the owner of the qutrit performs the first nontrivial
measurement has never been touched yet since the work of Walgate and
Hardy \cite{WH02}.

The purpose of this paper is to study the local distinguishability
of $2\otimes3$ states. We assume the dimension of Alice's system is
$2$, and the dimension of Bob's system is $3$. Due to the result in
Ref. \cite{WH02}, we only consider the case when Bob goes first,
which means that Bob first does a nontrivial measurement on his own
system. We find that for the discrimination of six states and five
states, LOCC and LPCC are equally powerful, i.e., a set of six or
five $2\otimes 3$ states is locally distinguishable if and only if
they are distinguishable by LPCC. But for four states and three
states, there exists a large class of states which can be
distinguished by LOCC, but not by LPCC. Therefore, we conclude that
local POVM is strictly powerful than local projective measurements
even for $2\otimes 3$ system. Furthermore, we obtain a complete
characterization of four $2\otimes 3$ states that are
distinguishable by LOCC but not by LPCC. For three states, such a
characterization is very difficult to obtain. Nevertheless, we
construct a general class of three states which are distinguishable
by LOCC but not by LPCC.  A feasible procedure for determining the
local distinguishability of three states is also presented.

We further study the effect of classical communication for
discrimination. We show that in general many rounds of classical
communication are necessary. We demonstrate this result by
constructing a class of $m\otimes n$ orthogonal states which
requires at least $2\min\{m,n\}-2$ rounds to achieve a perfect
discrimination. In some sense, our result is in accordance with the
recent result by Owari and Hayashi in Ref. \cite{OH07}, where they
showed that two-way classical communication can effectively increase
the local distinguishability. We would like to point out that the
problem studied in Ref. \cite{OH07} is quite different from ours.
More precisely, Ref. \cite{OH07} considers the discrimination
between a pure state and a mixed state, and requires the detection
of pure state can be achieved perfectly. The goal is to minimize the
minimal error of detecting the mixed state. Here we only consider
pure states and require each state to be identified perfectly.

The rest of the paper is organized as follows. In Section II we
first give a characterization for the distinguishability of
$2\otimes 3$ states by LPCC. Then in Section III and Section IV we
present in sequel our results about the local distinguishability of
six and five $2\otimes 3$ states. Section V and VI devote to the
local distinguishability of four and three states, respectively. In
Section VII we present a non-trivial set of bipartite pure states
which requires multi-round of classical communication to achieve a
perfect discrimination. We conclude the paper with a brief
discussion in Section VIII.

For simplicity,  in what follows we shall write
$\ket{\alpha}=\ket{\beta}$ for any two states which are different
from each other only with a nonzero factor. Sometimes we simply say
POVM or projective measurements instead of local POVM or local
projective measurements, respectively.
\section{Distinguishability of $2\otimes 3$ states by LPCC}
In Ref. \cite{BDM+99} Bennett and coworkers showed that any finite
set of $2\otimes n$ orthogonal product states can be perfectly
distinguishable using LPCC. For $2\otimes 3$ states this interesting
result has a converse as follows.
\begin{theorem}\upshape
A set of $2\otimes 3$ states are distinguishable by LPCC only if
there is a set of orthogonal product states such that each of the
given states can be written as a disjoint summation of these product
states.
\end{theorem}

Let us make the above theorem more transparent. Suppose
$\{\ket{\psi_k}:k=1,\cdots, n\}$ is a set of $2\otimes 3$ states.
Then these states are distinguishable by LPCC if and only if there
exists a set of orthogonal product states
$\{\ket{\phi_j}:j=1,\cdots, m\}$ and a partition of $\{1,\cdots,
m\}$, say $S_1,\cdots, S_n$, such that $\ket{\psi_k}\in
span\{\ket{\phi_j}: j\in S_k\}$, where $\cup_{i=1}^m
{S_i}=\{1,\cdots, m\}$ and $S_i\cap S_j=\emptyset$ for any $i\neq
j$.

{\bf Proof.} If Alice goes first, then it has been proven in Ref.
\cite{WH02} that a set of $2\otimes n$ states is distinguishable by
LOCC can be written as the sum of states from a set of orthogonal
product states.

Suppose now Bob goes first. If Bob's measurement operators can be
written as $P_1=\op{0}{0}+\op{1}{1}$ and $P_2=\ket{2}\bra{2}$,  then
if measurement result is 1, we can write Alice's measurement as
$\ket{0}\bra{0}$ and $\ket{1}\bra{1}$. We construct a set of
orthogonal product states as follows: $\{\ket{\alpha}_A\ket{2}_B$,
$\ket{\alpha^\bot}_A\ket{2}_B$, $\ket{0}_A\ket{\beta}_B$,
$\ket{0}_A\ket{\beta^\bot}_B$, $\ket{1}_A\ket{\gamma}_B$,
$\ket{1}_A\ket{\gamma^\bot}_B\}$, where $\ket{\beta}$,
$\ket{\beta^\bot}$, $\ket{\gamma}$, $\ket{\gamma^\bot}$ belong to
$span\{\ket{0},\ket{1}\}$. It is easy to see that if a set of states
is distinguishable by the LPCC we write above, they can be rewritten
as sum of states from the above set.

If Bob's measurement operators can be written as: $P_1=\op{0}{0}$,
$P_2=\op{1}{1}$ and $P_3=\op{2}{2}$. Then we construct a set of
orthogonal product states as: $\{\ket{\alpha}_A\ket{0}_B$,
$\ket{\alpha^\bot}_A\ket{0}_B$, $\ket{\beta}_A\ket{1}_B$,
$\ket{\beta^\bot}_A\ket{1}_B$, $\ket{\gamma}_A\ket{2}_B$,
$\ket{\gamma^\bot}_A\ket{2}_B\}$. Obviously, if a set of states can
be distinguished by the LPCC we write above, these states can be
rewritten as the disjoint sum of the states from the above set.
\hfill $\square$

\section{six states}

Six states form a complete basis for $2\otimes 3$ system. By a
result of Horodecki $et~al$ \cite{HSSH03}, we know that six states
are distinguishable by LOCC only if they are product states.
Conversely, by the result of Bennett $et~al$ mentioned above, we
conclude that any $2\otimes 3$ product basis are perfectly
distinguishable by LPCC. Thus we arrive at the following:
\begin{theorem}\label{six-states}\upshape
Six orthogonal $2\otimes 3$ states are perfectly distinguishable by
LPCC if and only if they form a complete orthogonal product basis.
Furthermore, the condition for the LOCC distinguishability is the
same for the LPCC distinguishability.
\end{theorem}

\section{five states}

\begin{theorem}\label{5-locc}\upshape
Five orthogonal $2\otimes3$ states are locally distinguishable if
and only if at most one of them is entangled.
\end{theorem}

{\bf Proof.} This result is a direct consequence of a more general
result presented in Ref. \cite{DFXY07}. Here we present a
self-contained proof.

Suppose the nontrivial measurement performed by Bob is $\{M_m\}$. We
consider $rank(M_m)$ and sort the condition according to the $M_m$'s
rank.

If $rank(\emph{M}_\emph{m})=3$, then on the next step, Alice should
distinguish $5$ orthogonal states
$I\otimes\emph{M}_\emph{m}\ket{\psi_i}$, which is possible only if
at most one of the five states is entangled. Because
$\emph{M}_\emph{m}$ is of full rank, it does not change the property
of being entangled or separable, so at most one of the five original
states is entangled.

If $rank(M_m)=2$, and $M_m\ket{\beta}=0$. Let $B'$ denote the
subspace orthogonal to $\ket{\beta}$, and the system $AB'$ is
$2\otimes2$. As Alice can distinguish at most $4$ orthogonal states
in $AB'$, one state must be eliminated after Bob's measurement,
denoted as $\ket{\psi_0}=\ket{\alpha}_A\ket{\beta}_B$. The other
four states are:
\begin{eqnarray}
\ket{\psi_1}=\ket{\eta_1}_{AB'}+\lambda_1\ket{\alpha_\bot}_A\ket{\beta},\nonumber\\
\ket{\psi_2}=\ket{\eta_2}_{AB'}+\lambda_2\ket{\alpha_\bot}_A\ket{\beta},\nonumber\\
\ket{\psi_3}=\ket{\eta_3}_{AB'}+\lambda_3\ket{\alpha_\bot}_A\ket{\beta},\nonumber\\
\ket{\psi_4}=\ket{\eta_4}_{AB'}+\lambda_4\ket{\alpha_\bot}_A\ket{\beta}.
\end{eqnarray}

If one of $\ket{\eta_i}$ is $0$, suppose it is $\ket{\eta_1}$, then
to keep orthogonality, $\lambda_1\neq 0$, and $\ket{\psi_1}$ is
product state, while the rest $\lambda_i=0$. As $I\otimes
M_m\ket{\eta_i}$ can be distinguished by Alice, at most one of
$I\otimes M_m\ket{\eta_i}$ is entangled state. On the system $B'$,
$M_m$ is of full rank, so at most one of the left three states
$\ket{\psi_i}=\ket{\eta_i}$ is entangled. We then reach the
conclusion that at most one state is entangled.

Suppose $\ket{\eta_i}\neq0$ for each $i$. As Alice uses projective
measurements $\ket{0}\bra{0}$ and $\ket{1}\bra{1}$ to distinguish
states after Bob's measurement, the five states can be rewritten as:
\begin{eqnarray}
\ket{\psi_0}=\ket{\alpha}_A\ket{\beta}_B,\\
\ket{\psi_1}=\ket{0}_A\ket{\gamma_1}_B+\lambda_1\ket{\alpha_\bot}_A\ket{\beta},\nonumber\\
\ket{\psi_2}=\ket{0}_A\ket{\gamma_2}_B+\lambda_2\ket{\alpha_\bot}_A\ket{\beta},\nonumber\\
\ket{\psi_3}=\ket{1}_A\ket{\gamma_3}_B+\lambda_3\ket{\alpha_\bot}_A\ket{\beta},\nonumber\\
\ket{\psi_4}=\ket{1}_A\ket{\gamma_4}_B+\lambda_4\ket{\alpha_\bot}_A\ket{\beta}.
\end{eqnarray}
To keep orthogonality relation $\ip{\psi_k}{\psi_l}=0$, where $k\in
\{1,2\}$ and $l\in \{3,4\}$, we have $\lambda_1=\lambda_2=0$ or
$\lambda_3=\lambda_4=0$. Suppose $\lambda_3=\lambda_4=0$,
$\lambda_1$ and $\lambda_2$ are not $0$, and
$\ip{\gamma_3}{\gamma_4}=0$.

If $\ket{\alpha}=\ket{1}$, then the five states are all product
states. We suppose $\ket{\alpha}\neq \ket{1}$. Then for an arbitrary
$E_m$, the following condition is satisfied:
$\bra{\gamma_1}E_m\ket{\beta}=\bra{\gamma_2}E_m\ket{\beta}=\bra{\gamma_3}E_m\ket{\gamma_4}=0$.

We choose $E_m$ satisfying $E_m\ket{\beta}\neq 0$. After Bob's
measurement, Alice does a projective measurement $\{\ket{0'},
\ket{1'}\}$. If $\{\ket{0'}, \ket{1'}\}$ = $\{\ket{\alpha},
\ket{\alpha^\bot}\}$, then as
$\bra{\psi_1}(\ket{\alpha}\bra{\alpha}\otimes E_m)\ket{\psi_2}=0$,
we have $\bra{\gamma_1}E_m\ket{\gamma_2}=0$. So
$\bra{\psi_1}(I\otimes
E_m)\ket{\psi_2}=\lambda_1\lambda_2^*\bra{\beta}E_m\ket{\beta}=0$,
one of $\lambda_1$ and $\lambda_2$ is $0$. If $\{\ket{0'},
\ket{1'}\}$ $\neq$ $\{\ket{\alpha}, \ket{\alpha^\bot}\}$, then after
Alice's measurement, at most three states are left as the dimension
of Bob's system is $3$. Because $E_m\ket{\beta}\neq 0$, the first
three states are not $0$. So the last two states are eliminated.
Then we have $E_m\ket{\gamma_3}=0$ and $E_m\ket{\gamma_4}=0$.
$\ket{\gamma_3}$ and $\ket{\gamma_4}$ form a basis of the subspace
orthogonal to $\ket{\beta}$, thus $E_m=k\ket{\beta}\bra{\beta}$. As
$\bra{\psi_2}I\otimes
E_m\ket{\psi_3}=\bra{\gamma_1}E_m\ket{\gamma_2}+\lambda_1\lambda_2^*\bra{\beta}E_m\ket{\beta}=k\lambda_1\lambda_2^*=0$,
that indicates one of $\lambda_1$ and $\lambda_2$ is also $0$. So
one of $\ket{\psi_2}$ and $\ket{\psi_3}$ is product state. Noticing
that we already have three product states: $\ket{\psi_1}$,
$\ket{\psi_4}$ and $\ket{\psi_5}$, we now have four product states.

If $rank(E_m)=1$ for each $m$, then
$E_m=\lambda_m\ket{\beta}\bra{\beta}$ for some $\lambda_m>0$. Let
$B'$ denote the subspace orthogonal to $\ket{\beta}$. Then five
states can be rewritten as
\begin{eqnarray}
\ket{\alpha}_A\ket{\beta}_B+\ket{\eta_0}_{AB'},\nonumber\\
\ket{\alpha^\bot}_A\ket{\beta}_B+\ket{\eta_1}_{AB'},\nonumber\\
\ket{\eta_2}_{AB'},~\ket{\eta_3}_{AB'},\ket{\eta_4}_{AB'}.
\end{eqnarray}
where $\ip{\eta_i}{\eta_j}=0$.

The system $AB'$ is $2\otimes2$, so at most four states can be
orthogonal to each other, then one of $\ket{\eta_0}_{AB'}$ and
$\ket{\eta_1}_{AB'}$ should vanish. Therefore, one of $\ket{\psi_0}$
and $\ket{\psi_1}$ is  a product state, and can be written as
$\ket{\alpha}_A\ket{\beta}_B$ corresponds to
$E_m=\lambda_m\ket{\beta}\bra{\beta}$. The number of measurement
operators is at least $3$ in order to satisfy $\sum E_m=I$. If the
number is $3$, then the measurement is actually a projective
measurement. By the result in Sec. II, at most one entangled state
exists. We only need to consider the case when the number is larger
than $3$. In this case we have at least $4$ different measurement
operators and each of them corresponds to a different product state.
So there must be at least $4$ product states.

From the discussion above, we conclude that at most one entangled
state exists in the set of five orthogonal states which are
distinguishable by LOCC.\hfill $\square$

From the above proof it is obvious that five states are perfectly
distinguished by LOCC if and only if they are perfectly
distinguished by LPCC.

\section{four states}

Now we consider the LOCC distinguishability of four states. We have
the following key theorem:
\begin{theorem}\label{schmidtsep}\upshape
Four orthogonal $2\otimes3$ states are perfectly distinguishable by
LOCC only if at least two of them are product states.
\end{theorem}

{\bf Proof.} We consider the rank of measurement operators performed
by Bob. There are three different cases:

Case 1: One of the measurement operators has rank $3$. After Bob's
measurement, none of the  states is eliminated. So Alice has to
distinguish $4$ orthogonal states, which is possible only when at
least two states are product states. As a full rank measurement
operator does not change the property of being entangled or
separable. Thus, at least two original states are product states.

Case 2: One of the measurement operator $M_1$ has rank $2$. Let us
assume $M_1\ket{2}_B=0$. There are three subcases we need to
consider:

Case 2.1: Two states can be written as $\ket{\alpha}\ket{2}$ and
$\ket{\beta}\ket{2}$. Then there are already two product states.

Case 2.2: Only one state can be written as $\ket{\alpha}\ket{2}$.
Then after Bob's measurement with outcome $1$, three states are
left. As Alice's measurement can only be a projective measurement of
the form $\{\ket{0}, \ket{1}\}$, we can rewrite three
post-measurement states as follows:
\begin{eqnarray}
\ket{0}_A\ket{\xi_1}_B+\ket{\eta_1}_A\ket{2}_B,\nonumber\\
\ket{1}_A\ket{\xi_2}_B+\ket{\eta_2}_A\ket{2}_B,\nonumber\\
\ket{0}_A\ket{\xi_3}_B+\ket{1}_A\ket{\xi_4}_B+\ket{\eta_3}_A\ket{2}_B.\nonumber
\end{eqnarray}
To keep orthogonality, we have
$\ip{\eta_1}{\eta_2}=\ip{\alpha}{\eta_2}=\ip{\alpha}{\eta_1}=0$. As
the dimension of Alice's system is $2$, from the equation above, we
have one of $\ket{\eta_1}$ and $\ket{\eta_2}$ is $0$. Then there are
two product states.

Case 2.3: No state can be written as $\ket{\alpha}\ket{2}$. The four
original states must be written as:
\begin{eqnarray}
\ket{0}_A\ket{\alpha_1}_B+\ket{\eta_1}_A\ket{2}_B,\nonumber\\
\ket{0}_A\ket{\alpha_2}_B+\ket{\eta_2}_A\ket{2}_B,\nonumber\\
\ket{1}_A\ket{\alpha_3}_B+\ket{\eta_3}_A\ket{2}_B,\nonumber\\
\ket{1}_A\ket{\alpha_4}_B+\ket{\eta_4}_A\ket{2}_B.\nonumber
\end{eqnarray}

We choose another measurement operator $M_2$ satisfying
$M_2\ket{2}\neq 0$, which always exists. If $M_2$'s condition can be
sorted into above cases 2.1 and 2.2, then we reach the conclusion
that two states are product states. So we only have to prove the
case that $I\otimes M_2\ket{\psi_i} \neq0$ for each $i$, which
suggests that if the result is $2$, all post-measurement states are
product states.

If $\ket{\eta_i}\neq 0$, then only under the condition that
$M_2\ket{\alpha_i}=\lambda_iM_2\ket{2}$ could $I\otimes
M_2\ket{\psi_i}$ be product state. The states after measurement can
be written as: $I\otimes
M_2\ket{\psi_i}=(\lambda_i\ket{0}+\ket{\eta_i})M\ket{2}$ or
$(\lambda_i\ket{1}+\ket{\eta_i})M\ket{2}$. Suppose at most one of
$\ket{\eta_i}$ is $0$, which means that there are at most one
product state. If one of $\ket{\eta_i}$ is $0$, then the state is
already product state. To keep orthogonality, one of the rest three
states satisfying $\lambda_i\ket{0}+\ket{\eta_i}$ or
$\lambda_i\ket{1}+\ket{\eta_i}$ is $0$, which means the state is
also product state. We then have two product states. If none of
$\ket{\eta_i}$ is $0$, then to keep orthogonality, two
$\lambda_i\ket{0}+\ket{\eta_i}$ or $\lambda_i\ket{1}+\ket{\eta_i}$
are $0$, and the two states are product states.

Case 3: Every measurement operator has rank $1$, and can be written
as $\emph{E}_\emph{i}=\ket{e_i}\bra{e_i}$ (unnormalized). Let
$\ket{\psi_i}=\ket{0}_A\ket{\alpha_i}+\ket{1}_A\ket{\beta_i}$. After
the measurement, two states must be eliminated to keep
orthogonality. It means for each $\ket{e_i}$, there are two states
$\ket{\psi_{i_1}}$ and $\ket{\psi_{i_2}}$ satisfying
$$(I\otimes\ket{e_i}\bra{e_i})\ket{\psi_{i_k}}=0, k=1,2,$$
where $i$ can take at least three different values as  there are at
least three measurement operators. So we have at least six
orthogonal equations. Then there are two states $\ket{\psi_i}$ such
that both of them have two orthogonality equations, i.e., each of
them is orthogonal to two $\ket{e_k}$. We can write these
orthogonality equations explicitly as follows:
$\ip{\alpha_i}{e_{i1}}=\ip{\alpha_i}{e_{i2}}=\ip{\beta_i}{e_{i1}}=\ip{\beta_i}{e_{i1}}=0$.
As $\ket{e_{i1}}$ and $\ket{e_{i2}}$ are linearly independent,
$\ket{\alpha_i}=\ket{\beta_i}$. It follows that two states should be
product states. \hfill $\square$

But different from the condition of five or six states, we find two
classes of four states which can only be distinguished by LOCC but
not by projective measurements. The result suggests that LOCC are
more powerful than LPCC. We list our results as the two following
theorems and each theorem discusses one class of states. We then
prove that these two classes consist of all states which can be
distinguished by LOCC but not by LPCC.

\begin{theorem}\label{4-special}\upshape
The following four orthogonal $2\otimes 3$ states can be
distinguishable by LOCC but not by LPCC.
\begin{eqnarray}
\ket{0}_A\ket{0}_B,~~~\ket{1}_A\ket{\alpha}_B,\nonumber\\
\ket{0}_A(a_1\ket{1}+b_1\ket{2})_B+\ket{1}_A(c_1\ket{\alpha^\bot}+d_1\ket{2})_B,\nonumber\\
\ket{0}_A(a_2\ket{1}+b_2\ket{2})_B+\ket{1}_A(c_2\ket{\alpha^\bot}+d_2\ket{2})_B,
\end{eqnarray}
where $a_1a_2^*+c_1c_2^*=b_1b_2^*+d_1d_2^*=0$, $\ket{\alpha}$ and
$\ket{\alpha^\bot}$ belong to $span\{\ket{0},\ket{1}\}$,
$\ket{\alpha}\neq\ket{0}$,
$k=-\frac{a_1a_2^*}{b_1b_2^*}=-\frac{c_1c_2^*}{d_1d_2^*}$ is real
number and satisfies $0<k<1$.
\end{theorem}

{\bf Proof.} First, to prove the states can be distinguished by
LOCC, we give a set of Bob's measurement operators:

\begin{eqnarray*}M_{1}=\left(
              \begin{array}{ccc}
                1 & 0 & 0 \\
                0 & 1 & 0 \\
                0 & 0 & \sqrt{k} \\
              \end{array}
            \right),
\  \   M_{2}=\left(
              \begin{array}{ccc}
                0 & 0 & 0 \\
                0 & 0 & 0 \\
                0 & 0 & \sqrt{1-k} \\
              \end{array}
            \right).\end{eqnarray*}

If the measurement outcome is $1$, then four post-measurement states
would be:
\begin{eqnarray}
\ket{0}_A\ket{0}_B, ~~\ket{1}_A\ket{\alpha}_B,\nonumber\\
\ket{0}_A(a_1\ket{1}+\sqrt{k}b_1\ket{2})_B+\ket{1}_A(c_1\ket{\alpha^\bot}+\sqrt{k}d_1\ket{2})_B,\nonumber\\
\ket{0}_A(a_2\ket{1}+\sqrt{k}b_2\ket{2})_B+\ket{1}_A(c_2\ket{\alpha^\bot}+\sqrt{k}d_2\ket{2})_B.
\end{eqnarray}

The above four states then can be distinguished by Alice with a
projective measurement $\{\ket{0},\ket{1}\}$.

If the measurement result is $2$, then two left states are:
\begin{eqnarray}
\sqrt{1-k}(b_1\ket{0}+d_1\ket{1})_A\ket{2}_B,\nonumber\\
\sqrt{1-k}(b_2\ket{0}+d_2\ket{0})_A\ket{2}_B.
\end{eqnarray}
We can verify that the above two states are orthogonal, and thus can
be perfectly distinguished. As a result, the original four states
can be perfectly distinguished by LOCC.

Next we shall show  that the above four states cannot be
distinguished by LPCC. Suppose Bob goes first. Since $\sum E_m=I$,
there is a rank one projective measurement operator which can be
written as: $P_1=\ket{\theta}\bra{\theta}$. If
$\ip{\theta}{0}\neq0$, then
$(I\otimes\ket{\theta}\bra{\theta})\ket{\psi_0}=\ip{\theta}{0}\ket{0}\ket{\theta}$.
To keep orthogonality between $\ket{\psi_0}$, $\ket{\psi_2}$, and
$\ket{\psi_3}$, $\ket{\theta}$ should be orthogonal to
$a_1\ket{1}+b_1\ket{2}$ and $a_2\ket{1}+b_2\ket{2}$. The above two
states are linear independent because if
$a_1\ket{1}+b_1\ket{2}=\lambda(a_2\ket{1}+b_2\ket{2}$), then
$k=-\frac{a_1a_2^*}{b_1b_2^*}=-\frac{a_1a_1^*}{b_1b_1^*}<0$. Then
$\ip{\theta}{1}=\ip{\theta}{2}=0$, $\ket{\theta}=\ket{0}$. However,
the projector $\ket{0}\bra{0}$ cannot keep orthogonality, because we
can prove that orthogonality require $k=1$. Thus the assumption of
$\ip{\theta}{0}\neq 0$ is incorrect. So we should have
$\ip{\theta}{0}=0$.

Similarly, we can prove $\ip{\theta}{\alpha}=0$. As
$\ket{\alpha}\neq\ket{0}$, $\ket{\theta}=\ket{2}$. So the rank one
projective measurement operator is $\ket{2}\bra{2}$. The left
projective measurement operator is
$P_2=\ket{0}\bra{0}+\ket{1}\bra{1}$. Unfortunately, by a similar
argument we can show that $P_2\ket{\psi_i}$ cannot be distinguished
by Alice. So the original four states cannot be distinguished by
projective measurements if Bob goes first.

If instead of Bob going first, Alice goes first, then after Alice's
measurement at most three states are left. In order to eliminate one
state, Alice's measurement operators must be $\ket{0}\bra{0}$ and
$\ket{1}\bra{1}$. But these operators cannot keep orthogonality
between the four states. So the original four states cannot be
distinguished by LOCC if Alice goes first. Thus we finish our proof.

An explicit example is as follows:
\begin{eqnarray}
\ket{0}_A\ket{0}_B,~~\ket{1}_A\ket{1}_B,\nonumber\\
\ket{0}_A(\ket{1}+\ket{2})_B+\ket{1}_A(\ket{0}-2\ket{2})_B,\nonumber\\
\ket{0}_A(\ket{1}-2\ket{2})_B-\ket{1}_A(\ket{1}+\ket{2})_B.
\end{eqnarray}
The measurement operators performed by Bob are:
\begin{eqnarray*}M_{1}=\left(
              \begin{array}{ccc}
                1 & 0 & 0 \\
                0 & 1 & 0 \\
                0 & 0 & \sqrt{1/2}\\
              \end{array}
            \right),
~M_{2}=\left(
              \begin{array}{ccc}
                0 & 0 & 0 \\
                0 & 0 & 0 \\
                0 & 0 & \sqrt{1/2}\\
              \end{array}
            \right).\end{eqnarray*}
Another different class of states is as follows:
\begin{theorem}\label{4-special2}\upshape
The following four orthogonal states can be distinguished by LOCC
but not by LPCC.
\begin{eqnarray}
\ket{0}_A\ket{0}_B,~~ \ket{\alpha}_A\ket{1}_B,\nonumber\\
a_1\ket{1}_A\ket{0}_B+b_1\ket{\alpha^\bot}_A\ket{1}_B+c_1\ket{1}_A\ket{2}_B,\nonumber\\
a_2\ket{1}_A\ket{0}_B+
b_2\ket{\alpha^\bot}_A\ket{1}_B+c_2\ket{\alpha^\bot}_A\ket{2}_B,
\end{eqnarray}
where $a_1a_2^*+b_1b_2^*+c_1c_2^*\ip{\alpha^\bot}{1}=0$, and
$k=-\frac{a_1a_2^*}{c_1c_2^*\ip{\alpha^\bot}{1}}$ is real number
which satisfies $0<k<1$. $\ket{\alpha}\neq\ket{0}$ and $\ket{1}$.
\end{theorem}

{\bf Proof.} Consider the following general measurement:
\begin{eqnarray*}M_{1}=\left(
              \begin{array}{ccc}
                1 & 0 & 0 \\
                0 & 0 & 0 \\
                0 & 0 & \sqrt{k} \\
              \end{array}
            \right),
\  \   M_{2}=\left(
              \begin{array}{ccc}
                0 & 0 & 0 \\
                0 & 1 & 0 \\
                0 & 0 & \sqrt{1-k} \\
              \end{array}
            \right).\end{eqnarray*}
If the measurement outcome is $1$, then after the measurement, three
left states are:
\begin{eqnarray}
\ket{0}_A\ket{0}_B,\nonumber\\
a_1\ket{1}_A\ket{0}_B+c_1\sqrt{k}\ket{1}_A\ket{2}_B,\nonumber\\
a_2\ket{1}_A\ket{0}_B+c_2\sqrt{k}\ket{\alpha^\bot}_A\ket{2}_B.
\end{eqnarray}
Alice can distinguish the states using projective measurements
$\ket{0}\bra{0}$ and $\ket{1}\bra{1}$. If the measurement result is
$2$, then three left states after the measurement are:
\begin{eqnarray}
\ket{\alpha}_A\ket{1}_B,\nonumber\\
b_1\ket{\alpha^\bot}_A\ket{1}_B+\sqrt{1-k}c_1\ket{1}_A\ket{2}_B,\nonumber\\
b_2\ket{\alpha^\bot}_A\ket{1}_B+\sqrt{1-k}c_2\ket{\alpha^\bot}_A\ket{2}_B.
\end{eqnarray}
Alice can distinguish the three states using
$\ket{\alpha}\bra{\alpha}$ and $\ket{\alpha^\bot}\bra{\alpha^\bot}$.

The next part is to prove that the four states cannot be
distinguished by LPCC. As $\sum P_m=I$, there is
$P_1=\ket{\theta}\bra{\theta}$. If $\ip{\theta}{0}\neq0$, then as
$\ip{\alpha}{0}\neq0$, to keep orthogonality between $\ket{\psi_1}$
and $\ket{\psi_2}$ we have $\ip{\theta}{1}=0$. Because
$\ip{\alpha^\bot}{0}\neq0$, to keep orthogonality between
$\ket{\psi_1}$ and $\ket{\psi_4}$ we also have $\ip{\theta}{2}=0$,
therefore $\ket{\theta}=\ket{0}$. But $\ket{0}\bra{0}$ cannot
distinguish the four states, so $\ip{\theta}{0}$ must be $0$.

Using the same method, we can also first assume $\ip{\theta}{1}\neq
0$, then we prove that $\ket{\theta}\bra{\theta}$ cannot distinguish
states, so $\ip{\theta}{1}= 0$. Therefore $\ket{\theta}=\ket{2}$.
But $\bra{\psi_3}(I\otimes\ket{2}\bra{2})\ket{\psi_4}\neq0$, so the
four states cannot be distinguished by projective measurements if
Bob goes first.

Suppose Alice goes first. After Alice does any operator
$\ket{\theta}\bra{\theta}$ there are at most three states left as
the dimension of Bob's system is $3$. As $\ket{\psi_3}$ and
$\ket{\psi_4}$ are entangled states,
$(I\otimes\ket{\theta}\bra{\theta})\ket{\psi_{3(4)}}\neq0$, so one
of $(I\otimes\ket{\theta}\bra{\theta})\ket{\psi_{1(2)}}=0$.
$\ket{\theta}$ must be $\ket{1}$ or $\ket{\alpha^\bot}$. But
$\ket{1}\bra{1}$ and $\ket{\alpha^\bot}\bra{\alpha^\bot}$ cannot
keep orthogonality. So the four states cannot be distinguished if
Alice goes first. Hence we finish the proof. \hfill $\square$

We also give an explicit example:
\begin{eqnarray}
\ket{0}_A\ket{0}_B,~~
(\frac{\ket{0}+\ket{1}}{\sqrt{2}})_A\ket{1}_B,\nonumber\\
-\frac{1}{2}\ket{1}_A\ket{0}_B+(\frac{\ket{0}-\ket{1}}{2\sqrt{2}})_A\ket{1}_B+\ket{1}_A\ket{2}_B,\nonumber\\
\frac{1}{\sqrt{2}}\ket{1}_A\ket{0}_B-(\frac{\ket{0}-\ket{1}}{2})_A\ket{1}_B-(\frac{\ket{0}-\ket{1}}{\sqrt{2}})_A\ket{2}_B.\nonumber
\end{eqnarray}
The measurement performed by Bob is given by
\begin{eqnarray*}M_{1}=\left(
              \begin{array}{ccc}
                1 & 0 & 0 \\
                0 & 0 & 0 \\
                0 & 0 & \sqrt{1/2} \\
              \end{array}
            \right),
\  \   M_{2}=\left(
              \begin{array}{ccc}
                0 & 0 & 0 \\
                0 & 1 & 0 \\
                0 & 0 & \sqrt{1/2} \\
              \end{array}
            \right).\end{eqnarray*}

Interestingly, the above two classes of states completely
characterize the local distinguishability of four $2\otimes 3$
states.
\begin{theorem}\label{4-special3}\upshape Any four $2\otimes 3$ orthogonal states can be distinguished by
LOCC but not by LPCC if and only if they can be written as one of
the form in the above two theorems.
\end{theorem}

{\bf Proof.} We have proved in Lemma 2 that two of the four states
should be product states if they are distinguishable by LOCC.

Case 1: If there are two states that can be written as
$\ket{0}_A\ket{0}_B$ and $\ket{1}_A\ket{\eta_0}_B$, where
$\ket{\eta_0}$ and $\ket{\eta_0^\bot}$ belong to
$span\{\ket{0},\ket{1}\}$. Then the other two states can be written
as:
\begin{eqnarray}
\ket{0}_A\ket{\alpha_1}_B+\ket{1}_A\ket{\eta_1}_B,\nonumber\\
\ket{0}_A\ket{\alpha_2}_B+\ket{1}_A\ket{\eta_2}_B,
\end{eqnarray}
where $\ket{\alpha_1}$ and $\ket{\alpha_2}$ belong to
$span\{\ket{1},\ket{2}\}$, $\ket{\eta_1}$ and $\ket{\eta_2}$ belong
to $span\{\ket{\eta_0^\bot},\ket{2}\}$.

We assume that $\ket{\alpha_1}\neq\ket{\alpha_2}$ and
$\ket{\eta_1}\neq\ket{\eta_2}$ and $\ket{\eta_0}\neq\ket{0}$. Other
cases such as as $\ket{\alpha_1}=\lambda\ket{\alpha_2}$ or
$\ket{\eta_1}=\lambda\ket{\eta_2}$  or $\ket{\eta_0}=\ket{0}$ will
be discussed later. To keep orthogonality after measurement,
$\bra{0}E_m\ket{\alpha_1}=\bra{0}E_m\ket{\alpha_2}=0$, as
$\ket{\alpha_1}\neq\ket{\alpha_2}$, we have
$\bra{0}E_m\ket{1}=\bra{0}E_m\ket{2}=0$. For the same reason,
$\bra{\eta_0}E_m\ket{\eta_0^\bot}=\bra{\eta_0}E_m\ket{2}=0$, as
$\ket{\eta_0}\neq\ket{0}$, $\bra{1}E_m\ket{2}=0$. We obtain that
$E_m$ is diagonal under the bases $\{\ket{0},\ket{1},\ket{2}\}$,
$E_m=diag(\lambda_0,\lambda_1,\lambda_2)$. We rewrite $\ket{\psi_3}$
and $\ket{\psi_4}$ as:
\begin{eqnarray}
\ket{0}_A(a_1\ket{1}+a_2\ket{2})_B+\ket{1}_A(a_3\ket{0}+a_4\ket{1}+a_5\ket{2})_B,\nonumber\\
\ket{0}_A(b_1\ket{1}+b_2\ket{2})_B+\ket{1}_A(b_3\ket{0}+b_4\ket{1}+b_5\ket{2})_B.\nonumber
\end{eqnarray}
To keep orthogonality, we should have $\braket{\psi_3}{I\otimes
E_m}{\psi_4}=0$, which is equivalent to
$$a_3b_3^*\lambda_0+(a_1b_1^*+a_4b_4^*)\lambda_1+(a_2b_2^*+a_5b_5^*)\lambda_2=0.$$
There are only two linearly independent solutions to the above
equation. Suppose $E_1=diag(\lambda_0, \lambda_1,\lambda_2)$ and
$E_2=diag(\lambda_0',\lambda_1',\lambda_2')$ are two independent
solutions. If we have another operator $E_3$, it must be written as
$E_3=aE_1+bE_2$, so we can use only $E_1$ and $E_2$ to distinguish
the states instead of using three or more operators. Therefore,
there are only two measurement operators $E_1=diag(\lambda_0,
\lambda_1,\lambda_2)$ and
$E_2=diag(1-\lambda_0,1-\lambda_1,1-\lambda_2)$.

If $\lambda_0\neq0$, $E_1\ket{0}\neq0$. After Alice's measurement
$\ket{\theta}\bra{\theta}$, $\bra{\psi_1}(\op{\theta}{\theta}\otimes
E_1)\ket{\psi_3}=\lambda_0\ip{0}{\theta}\ip{\theta}{1}\ip{0}{\eta_1}=0$.
If $\ket{\theta}\neq\ket{0}$ or $\ket{1}$, then as $\ket{\eta_1}$
belongs to $span\{\ket{\eta_0^\bot},\ket{2}\}$,
$\ket{\eta_1}=\ket{2}$. For the same reason
$\ket{\eta_2}=\ket{\eta_1}=\ket{2}$, but we have assumed in the
beginning that $\ket{\eta_2}\neq\ket{\eta_1}$. So
$\ket{\theta}=\ket{0}$ or $\ket{1}$ and Alice's measurement
operators are : $\ket{0}\bra{0}$ and $\ket{1}\bra{1}$. Similarly, if
If $\lambda_1\neq0$, then Alice's measurement operators must also be
$\ket{0}\bra{0}$ and $\ket{1}\bra{1}$.

From $\bra{\psi_3}(\ket{0}\bra{0}\otimes E_m)\ket{\psi_4}=0$, and
$\ip{\psi_3}{1}\bra{1}\otimes E_m\ket{\psi_4}=0$, we have
$a_1b_1^*\lambda_2+a_2b_2^*\lambda_3=0$ and
$a_3b_3^*\lambda_1+a_4b_4^*\lambda_2+a_5b_5^*\lambda_3=0$. If
$1-\lambda_1\neq0$ or $1-\lambda_2\neq0$, we also have
$a_1b_1^*(1-\lambda_2)+a_2b_2^*(1-\lambda_3)=0$ and
$a_3b_3^*(1-\lambda_1)+a_4b_4^*(1-\lambda_2)+a_5b_5^*(1-\lambda_3)=0$.
Then from those equations above, $a_1b_1^*+a_2b_2^*=0$ and
$a_3b_3^*+a_4b_4^*+a_5b_5^*=0$ stand, therefore the four states can
be distinguished by Alice first doing measurements $\ket{0}\bra{0}$
and $\ket{1}\bra{1}$. As we have assumed these four states cannot be
distinguished by projective measurements, we have either
$\lambda_1=\lambda_2=0$ or $1-\lambda_1=1-\lambda_2=0$. So the POVM
consists of $E_1=diag(1,1,k)$ and $E_2=diag(0,0,1-k)$ which is just
the case in theorem $4$.

Here we will discuss other conditions we mentioned in the beginning.
First, if $\ket{\eta_0}=\ket{0}$, then $\ket{0}\bra{0}$ can
distinguish $\ket{\psi_1}$ and $\ket{\psi_2}$. As $\ket{\alpha_i}$
and $\ket{\eta_i}$ belong to $span\{\ket{1},\ket{2}\}$,
$\ket{1}\bra{1}+\ket{2}\bra{2}$ can distinguish $\ket{\psi_3}$ and
$\ket{\psi_4}$. Therefore, the four states can be distinguished by
projective measurements.

Secondly, if $\ket{\eta_2}=\lambda\ket{\eta_1}$, we choose $M_1$
which satisfies $M_1\ket{\eta_1}\neq0$. Then Alice's measurement
cannot be $\ket{0}\bra{0}$ and $\ket{1}\bra{1}$, because
$\ket{\psi_3}(\ket{1}\bra{1}\otimes
E_1)\ket{\psi_4}=\lambda\bra{\eta_1}E_1\ket{\eta_1}\neq0$. Then to
keep orthogonality after measurements $\ket{0'}\bra{0'}\otimes M_1$
and $\ket{1'}\bra{1'}\otimes M_1$, we have
$\ket{\psi_1}(\ket{0'}\bra{0'}\otimes
E_1)\ket{\psi_2}=\ip{0}{0'}\ip{0'}{1}\bra{0}E_1\ket{\eta_0}=0$, so
$E_1\ket{0}\bot\ket{\eta_0}$. Similarly, we have $E_1\ket{0}$ $\bot$
$\{\ket{\eta_0},\ket{\alpha_1},\ket{\alpha_2},\ket{\eta_1}\}$ and
$E_1\ket{\eta_0}$ $\bot$
$\{\ket{0},\ket{\alpha_1},\ket{\alpha_2},\ket{\eta_1}\}$.

As $\ket{\eta_0}=a_1\ket{0}+a_2\ket{1}$,
$\ket{\alpha_i}=b_1\ket{1}+b_2\ket{2}$ and
$\ket{\eta_1}=c_1\ket{\eta_0^\bot}+c_2\ket{2}$, at most one of the
next two equations stands:
$\ket{\alpha_i}=\lambda_i\ket{\eta_0}+\mu_i\ket{\eta_1}$ and
$\ket{\alpha_i}=\lambda_i\ket{0}+\mu_i\ket{\eta_1}$.

If both of them does not stand, then the dimension of each set is 3.
So $E_1\ket{0}=E_1\ket{1}=0$, $E_1=\ket{2}\bra{2}$, and the four
states can be distinguishable by Bob's projective measurements
$\ket{0}\bra{0}+\ket{1}\bra{1}$ and $\ket{2}\bra{2}$.

Without loss of generality, we suppose the second one does not
stand, then we have $E_1\ket{\eta_0}=0$, and the four states can be
rewritten as:
\begin{eqnarray}
\ket{0}\ket{0},~\ket{1}\ket{\eta_0},\nonumber\\
\lambda_1\ket{0}\ket{\eta_0}+\ket{\beta_1}\ket{\eta_1},\nonumber\\
\lambda_2\ket{0}\ket{\eta_0}+\ket{\beta_2}\ket{\eta_1}.
\end{eqnarray}
$\bra{\psi_3}(I\otimes
E_1)\ket{\psi_4}=\lambda_1\lambda_2\bra{\eta_0}
E_1\ket{\eta_0}+\ip{\beta_1}{\beta_2}\bra{\eta_1} E_1\ket{\eta_1}=0
$, so $\ip{\beta_1}{\beta_2}=0$, then
$\ip{\psi_3}{\psi_4}=\lambda_1\lambda_2=0$. One of $\lambda_1$ and
$\lambda_2$ is 0, then the four states can be distinguished by
projective measurements. The condition
$\ket{\alpha_2}=\lambda\ket{\alpha_1}$ can be discussed similarly.

Case 2: If two product states can be written as $\ket{0}_A\ket{0}_B$
and $\ket{\alpha}_A\ket{1}_B$, then the other two are
\begin{eqnarray}
a_1\ket{1}_A\ket{0}_B+b_1\ket{\alpha^\perp}_A\ket{1}_B+\ket{\theta_1}_A\ket{2}_B,\nonumber\\
a_2\ket{1}_A\ket{0}_B+b_2\ket{\alpha^\perp}_A\ket{1}_B+\ket{\theta_2}_A\ket{2}_B.
\end{eqnarray}

As the condition that $\ket{\alpha}=\ket{1}$ can be count into case
$1$, we suppose here $\ket{\alpha}\neq\ket{1}$. If one of
$\ket{\theta_i}$ is $0$, then the four states can only be
distinguished by projective measurements $\ket{2}\bra{2}$ and
$\ket{0}\bra{0}+\ket{1}\bra{1}$. So we suppose none of
$\ket{\theta_i}$ is $0$. We also assume that at least one of $a_i$,
or  $b_i$ is not $0$, because otherwise the four states can be
distinguished by projective measurements.

To keep orthogonality between $\ket{\psi_1}$ and $\ket{\psi_2}$
after Bob's measurement, $\bra{0}E_m\ket{1}=0$. And as
$\bra{\psi_1}I\otimes
E_m\ket{\psi_3}=\ip{0}{\theta_1}\bra{0}E_m\ket{2}=0$ and
$\bra{\psi_1}I\otimes
E_m\ket{\psi_3}=\ip{0}{\theta_2}\bra{0}E_m\ket{2}=0$. If
$\ket{\theta_1}$ or $\ket{\theta_2}$ is not $\ket{1}$, then
$\bra{0}E_m\ket{2}=0$. Similarly, if $\ket{\theta_1}$ or
$\ket{\theta_2}$ is not $\ket{\alpha^\bot}$, then
$\bra{1}E_m\ket{2}=0$. So $E_m$ is diagonal,
$E_m=(\lambda_1,\lambda_2,\lambda_3)$.

We also suppose $\ket{\alpha}\neq\ket{0}$. The conditions such as
$\ket{\theta_1}=\ket{\theta_2}=\ket{1}$ or $\ket{\alpha^\bot}$ and
$\ket{\alpha}=\ket{0}$ will be discussed later. We choose $M_1$
satisfying $M_1\ket{0}\neq0$, and denote Alice's measurement
operators as $\ket{0'}\bra{0'}$ and $\ket{1'}\bra{1'}$. As we
suppose one of $a_i$ is not $0$, without losing generality,
$a_1\neq0$, then $\bra{\psi_1}(\ket{0'}\bra{0'}\otimes
E_1)\ket{\psi_2}=\ip{0}{0'}\ip{0'}{1}\bra{0}E_1\ket{0}=0$.
$\bra{0}E_1\ket{0}\neq0$, so either $\ip{0}{0'}=0$ or
$\ip{0'}{1}=0$, then Alice's measurements should be $\ket{0}\bra{0}$
and $\ket{1}\bra{1}$.

Similarly, we consider the distinguishability between $\ket{\psi_2}$
$\ket{\psi_3}$ and $\ket{\psi_4}$. If $M_1\ket{1}$ is also not $0$,
then Alice's measurements should be $\ket{\alpha}\bra{\alpha}$ and
$\ket{\alpha^\bot}\bra{\alpha^\bot}$. Notice that the two sets
$\{\ket{0}\bra{0},\ket{1}\bra{1}\}$ and
$\{\ket{\alpha}\bra{\alpha},\ket{\alpha^\bot}\bra{\alpha^\bot}\}$
are different as we suppose $\ket{\alpha}$ is not equal to $\ket{0}$
or $\ket{1}$, Alice cannot distinguish the four states after Bob's
measurement. Therefore, only one of $M_1\ket{0}$ and $M_1\ket{1}$ is
not $0$. As $E_m$ is diagonal , there are at most two linear
independent solutions of $(\lambda_1,\lambda_2,\lambda_3)$ which
results from similar discussion as in case $1$. The two measurements
can be written as: $E_1=diag(1,0,k)$ and $E_1=diag(0,1,1-k)$. If the
result is $1$, then Alice's measurements should be $\ket{0}\bra{0}$
and $\ket{1}\bra{1}$. $\bra{\psi_3}(\ket{0}\bra{0}\otimes
E_1)\ket{\psi_4}=k\ip{\theta_1}{0}\ip{\theta_2}{0}=0$, so one of
$\ket{\theta_i}$ is $\ket{1}$. For the same reason, the other is
$\ket{\alpha^\bot}$. It is the case in theorem $5$.

We discuss other conditions here. First, if
$\ket{\theta_1}=\ket{\theta_2}=\ket{1}$, then $\ket{\psi_3}$ and
$\ket{\psi_4}$ are:
\begin{eqnarray}
\ket{1}_A(a_1\ket{0}+b_1\ket{2})_B+c_1\ket{\alpha^\bot}_A\ket{1}_B,\nonumber\\
\ket{1}_A(a_2\ket{0}+b_2\ket{2})_B+c_2\ket{\alpha^\bot}_A\ket{1}_B.
\end{eqnarray}
If here $\ket{\alpha}=\ket{0}$, then the four states are all product
states, and can be distinguished by projective measurements. We then
suppose here $\ket{\alpha}\neq\ket{0}$. To keep orthogonality
between $M_m\ket{\psi_1}$ and $M_m\ket{\psi_2}$,
$\bra{0}E_m\ket{1}=0$. $\bra{\psi_2}(I\otimes
E_m)\ket{\psi_3}=b_1\ip{\alpha}{1}\bra{1}E_m\ket{2}=0$. As at least
one of $b_i$ is not 0, we have $\bra{1}E_m\ket{2}=0$.

We choose $M_1$ satisfying $M_1\ket{1}\neq0$. After Bob's
measurement, Alice should distinguish four states $I\otimes
M_1\ket{\psi_i}$. Suppose one of Alice's measurement operators is
$\ket{0'}\bra{0'}$, then $\bra{\psi_2}(\ket{0'}\bra{0'}\otimes
E_1)\ket{\psi_3}=\ip{\alpha}{0'}\ip{0'}{\alpha^\bot}^*b_1\bra{1}E_1\ket{1}=0$.
The equation is also satisfied for $b_2$. As we suppose one of $b_i$
is not $0$, $\ip{0'}{\alpha}=0$ or $\ip{0'}{\alpha^\bot}=0$. So
Alice's measurement should be $\ket{\alpha}\bra{\alpha}$ and
$\ket{\alpha^\bot}\bra{\alpha^\bot}$. As
$\bra{1}E_1\ket{0}=\bra{1}E_1\ket{2}=0$,
$\bra{\psi_3}(\ket{\alpha}\bra{\alpha}\otimes
E_1)\ket{\psi_4}=\ip{1}{\alpha}\ip{1}{\alpha}^*(a_1\bra{0}+c_1\bra{2})E_1(a_2\ket{0}+c_2\ket{2})=0$
, therefore $(a_1\bra{0}+c_1\bra{2})E_1(a_2\ket{0}+c_2\ket{2})=0$.
It results in
$\bra{\psi_3}(\ket{\alpha^\bot}\bra{\alpha^\bot}\otimes
E_1)\ket{\psi_4}=\ip{1}{\alpha^\bot}\ip{1}{\alpha^\bot}^*(a_1\bra{0}+c_1\bra{2})E_1(a_2\ket{0}+c_2\ket{2})+b_1b_2^*\bra{1}E_1\ket{1}=b_1b_2^*\bra{1}E_1\ket{1}=0$,
so one of $b_i$ is $0$. Then the four states can be distinguished by
Bob's measurement operators $\ket{1}\bra{1}$ and
$\ket{0}\bra{0}+\ket{2}\bra{2}$. Similarly, the condition that
$\ket{\theta_1}=\ket{\theta_2}=\ket{\alpha^\bot}$ can be discussed
using the above method.

Secondly, we discuss the condition that $\ket{\alpha}=\ket{0}$ while
one of $\ket{\theta_i}$ is not $\ket{1}$. The four states are:
\begin{eqnarray}
\ket{0}_A\ket{0}_B,~~~\ket{0}_A\ket{1}_B,\nonumber\\
\ket{1}_A(a_1\ket{0}+b_1\ket{1})_B+\ket{\theta_1}_A\ket{2}_B,\nonumber\\
\ket{1}_A(a_2\ket{0}+b_2\ket{1})_B+\ket{\theta_2}_A\ket{2}_B.
\end{eqnarray}

If one of $\ket{\theta_i}=\ket{1}$, we will have three product
states, then the four states can be distinguished by Alice first
doing measurement $\ket{0}\bra{0}$ and $\ket{1}\bra{1}$, so we
suppose none of $\ket{\theta_i}$ is equal to $\ket{1}$. From
orthogonality, we can get $E_m$ is diagonal. We choose $M_1$
satisfying $M_1\ket{0}\neq0$ or $M_1\ket{1}\neq0$, then as we proved
above, Alice's measurement should be $\ket{0}\bra{0}$ and
$\ket{1}\bra{1}$. As $\bra{\psi_3}(\ket{0}\bra{0}\otimes
E_1)\ket{\psi_4}=\ip{\theta_1}{0}\ip{\theta_2}{0}\bra{2}E_1\ket{2}=0$,
if $M_1\ket{0}\neq0$ or $M_1\ket{1}\neq0$, then $M_1\ket{2}=0$. As
$\sum E_m=I$, one of the measurement operator must be
$\ket{2}\bra{2}$, so $\ip{\theta_1}{\theta_2}=0$. The four states
can be distinguished by Bob's measurement $\ket{2}\bra{2}$ and
$\ket{0}\bra{0}+\ket{1}\bra{1}$. \hfill $\square$
\section{three states}

We can easily construct a class of three states that can be exactly
distinguishable by LOCC but not by LPCC as follows:
\begin{theorem}\label{3-special}\upshape
Three orthogonal $2\otimes3$ states
$\ket{\psi_i}=\ket{0}_A\ket{\eta_i}_B+\ket{1}_A\ket{\xi_i}_B$, which
have the following forms can be distinguishable by LOCC but not by
LPCC:
\begin{eqnarray}
\ket{0}_A\ket{0}_B,\nonumber\\
\ket{0}_A(a_1\ket{1}+a_2\ket{2})_B+\ket{1}_A(a_3\ket{0}+a_4\ket{1}+a_5\ket{2})_B,\nonumber\\
\ket{0}_A(b_1\ket{1}+b_2\ket{2})_B+\ket{1}_A(b_3\ket{0}+b_4\ket{1}+b_5\ket{2})_B,\nonumber
\end{eqnarray}
where $a_i$ and $b_i$ satisfy
$a_1b_1^*+a_2b_2^*+a_3b_3^*+a_4b_4^*+a_5b_5^*=0$, $\alpha a_1b_1^*+
\beta a_2b_2^*=0$, $a_3b_3^*+\alpha a_4b_4^*+ \beta a_5b_5^*=0$,
$\ip{\eta_2}{\eta_3}\neq0$, $\ $ $\ip{\xi_2}{\xi_3}\neq0$, $\ $
$\ip{\eta_2}{\eta_3}+\ip{\xi_2}{\eta_2}\ip{\eta_2}{\xi_3}\neq0$,
$\ip{\eta_2}{\eta_3}+\ip{\xi_2}{\eta_3}\ip{\eta_3}{\xi_3}\neq0$,
$\ket{\eta_2}\neq\ket{\eta_3}$, $a_3b_3^*\neq0$ and
$0<\alpha,\beta<1$.
\end{theorem}

{\bf Proof.}The POVM consists of two parts:
\begin{eqnarray*}M_{1}=\left(
              \begin{array}{ccc}
                1 & 0 & 0 \\
                0 & \sqrt{\alpha} & 0 \\
                0 & 0 & \sqrt{\beta}\\
              \end{array}
            \right),~
               M_{2}=\left(
              \begin{array}{ccc}
                0 & 0 & 0 \\
                0 & \sqrt{1-\alpha} & 0 \\
                0 & 0 & \sqrt{1-\beta}\\
              \end{array}
            \right).\end{eqnarray*}

If the measurement result is $1$, then after Bob's measurement the
three states are:
\begin{eqnarray}
\ket{0}\ket{0},\nonumber\\
a_3\ket{1}\ket{0}+(a_1\ket{0}+a_4\ket{1})\sqrt{\alpha}\ket{1}+(a_2\ket{0}+a_5\ket{1})\sqrt{\beta}\ket{2},\nonumber\\
b_3\ket{1}\ket{0}+(b_1\ket{0}+b_4\ket{1})\sqrt{\alpha}\ket{1}+(b_2\ket{0}+b_5\ket{1})\sqrt{\beta}\ket{2},\nonumber\\
\end{eqnarray}

Because of the relationship given above, three (unnormalized) states
$I\otimes M_1\ket{\psi_i}$ are orthogonal to each other and can be
distinguished if Alice performs a measurement $\ket{0}\bra{0}$ and $
\ket{1}\bra{1}$.

If measurement result is $2$, then there are only two orthogonal
states $I\otimes M_2\ket{\psi_2}$ and $I\otimes M_2\ket{\psi_3}$
left, and $\bra{\psi_2}I\otimes E_{2}\ket{\psi_3}=0$. So the two
states can be distinguished by LOCC.

The next part of the proof is to prove the three states cannot be
distinguished by projective measurements. Let
$P_1=\ket{\theta}\bra{\theta}$. We assume $\ip{\theta}{0}\neq0$,
then, to keep orthogonality between the three states, one state
should be eliminated if the measurement result is $1$. Without
losing generality, we can suppose $I\otimes P_1\ket{\psi_3}=0$, then
$\ip{\theta}{\eta_3}=\ip{\theta}{\xi_3}=0$. From
$\bra{\psi_1}I\otimes P_1\ket{\psi_2}=0$, we have
$\ip{\theta}{\eta_2}=0$. The conditions in the theorem indicates
that $\ket{\eta_2}$, $\ket{\eta_3}$, $\ket{\xi_3}$ are linear
independent, therefore $\ket{\theta}$ does not exist. Then
$\ip{\theta}{0}$ must be equal to $0$.

The left projective measurement is
$P_2=\ket{0}\bra{0}+\ket{\theta^\perp}\bra{\theta^\perp}$, where
$\ket{\theta^\perp}$ belongs to $span(\ket{1},\ket{2})$. Notice that
the necessary condition for Alice to distinguish three states is at
most one state is entangled, then one of
$I\otimes(\ket{0}\bra{0}+\ket{\theta^\perp}\bra{\theta^\perp})\ket{\psi_{2(3)}}$
must be product state. As $P_2\ket{\eta_i}\neq P_2\ket{\xi_i}$ and
$P_2\ket{\xi_i}\neq0$, we have $P_2\ket{\eta_i}=0$ if the state is
product state. It indicates that one of $\ket{\eta_2}$ or
$\ket{\eta_3}$ must be orthogonal to $\ket{\theta^\perp}$. Suppose
$\ket{\theta^\perp}$ is orthogonal to $\ket{\eta_2}$, then
$\ket{\eta_2}=\ket{\theta}$. Because the condition
$\ip{\eta_2}{\eta_3}+\ip{\xi_2}{\eta_2}\ip{\eta_2}{\xi_3}\neq0$ is
satisfied, $P_1=\ket{\theta}\bra{\theta}=\ket{\eta_2}\bra{\eta_2}$
cannot keep orthogonality between $\ket{\psi_2}$ and $\ket{\psi_3}$.
Thus, if Bob goes first, these states cannot be distinguished by
LPCC.

On the other hand, if Alice goes first, suppose Alice's measurement
is: $\ket{0'}\bra{0'}$ and $\ket{1'}\bra{1'}$. As
$\bra{\psi_1}(\op{0'}{0'}\otimes
I\ket{\psi_2}=a_3\ip{0}{0'}\ip{0'}{1}=0$, Alice's measurement must
be: $\ket{0}\bra{0}$ and $\ket{1}\bra{1}$. Because
$\ip{\eta_2}{\eta_3}\neq0$, Alice's measurement $\ket{0}\bra{0}$
cannot distinguish the four states. Therefore, the three states
cannot be distinguished by LPCC. Here, we finish our proof.

We give a specific example of three states which have the form in
the theorem:
\begin{eqnarray}
\ket{0}_A\ket{0}_B,\nonumber\\
\ket{0}_A(3\ket{0}+3\ket{2})_B+\ket{0}_A(\ket{0}+3\ket{1}-2\ket{2})_B,\nonumber\\
\ket{0}_A(3\ket{0}-2\ket{2})_B+\ket{0}_A(2\ket{0}-1\ket{1}+\ket{2})_B.
\end{eqnarray}

The POVM performed by Bob is as follows:
\begin{eqnarray*}M_{1}=\left(
              \begin{array}{ccc}
                1 & 0 & 0 \\
                0 & \sqrt{1/3}& 0 \\
                0 & 0 & \sqrt{1/2}\\
              \end{array}
            \right),
           ~ M_{2}=\left(
              \begin{array}{ccc}
                0 & 0 & 0 \\
                0 & \sqrt{2/3} & 0 \\
                0 & 0 & \sqrt{1/2}\\
              \end{array}
            \right).\end{eqnarray*}

It is easy to prove that the above three states can be distinguished
by the above POVM but not by projective measurements.

For three states, to determine whether they can be distinguished by
LOCC is much harder. We will give a protocol to determine whether
three given orthogonal states can be distinguished.

First, the three states $\ket{\psi_i}$ are denoted as:
$\ket{0}_A\ket{\alpha_i}_B+\ket{1}_A\ket{\beta_i}_B$. After the
measurement, the states are $I\otimes M\ket{\psi_i}$. Taking the
condition for Alice to distinguish three states into consideration ,
the three states after Bob's measurement can be written as:
$\ket{0^*}_A\ket{\xi_i}_B+\ket{1^*}_A\ket{\theta_i}_B$, where
$\ket{\xi_i}$ and $\ket{\theta_i}$ are two sets of orthogonal states
of Bob's system, $\ket{0^*}$ and $\ket{1^*}$ are two specific bases
of Alice's. In spite of coefficients, we have
$\sum\ket{\xi_i}\bra{\xi_i}=I$ and
$\sum\ket{\theta_i}\bra{\theta_i}=I$.

Suppose $\ket{0^*}=a\ket{0}+b\ket{1}$ and
$\ket{1^*}=-b^*\ket{0}+a^*\ket{1}$. Then we have
$\ket{0^*}\bra{0^*}\otimes
M_m\ket{\psi_i}=\ket{0^*}_AM_m(a\ket{\alpha_i}+b\ket{\beta_i})_B=\ket{0^*}_A\ket{\xi_i}_B$
and $\ket{1^*}\bra{1^*}\otimes
M_m\ket{\psi_i}=\ket{1^*}_AM_m(-b^*\ket{\alpha_i}+a^*\ket{\beta_i})_B=\ket{1^*}_A\ket{\theta_i}_B$.

Let $\ket{\phi_i}$ denote $a\ket{\alpha_i}+b\ket{\beta_i}$, then we
can construct another set of states $\ket{\eta_i}$ satisfying
$\ip{\eta_i}{\phi_j}=0$ for any $j\neq i$. Because $\ket{\xi_i}$ is
a set of orthogonal states, $\ip{\xi_i}{\xi_j}=\bra{\xi_i}
M\ket{\phi_j}=0$. Comparing to the definition of $\ket{\eta_i}$, we
have $\ket{\eta_i}= M\ket{\xi_i}$. So we can choose positive numbers
$\lambda_i$, to have the following equation satisfied:
$\sum{\lambda_i}\ket{\eta_i}\bra{\eta_i}=\sum{M^\dagger\ket{\xi_i}\bra{\xi_i}M}=M^\dag
M=E_m$.

If we let $\ket{\varphi_i}=-b^*\ket{\xi_i}+a^*\ket{\theta_i}$, then
using the same method, we can find $\ip{\mu_i}{\varphi_j}=0$, for
any $j\neq i$. We can also choose proper positive numbers $\nu_i$ to
have the following equation satisfied:
$\sum{\nu_i}\ket{\mu_i}\bra{\mu_i}=E_m$.

We finally have the equation
$$\sum{\nu_i}\ket{\mu_i}\bra{\mu_i}=\sum{\lambda_i}\ket{\eta_i}\bra{\eta_i}=E_m.$$
There are eight independent variables $a$, $b$, $\lambda_i$, $\nu_i$
and nine equations. Getting value of the variables which satisfying
the above equation, we can construct a set of POVM to distinguish
the three given states. From the equation, we can see that it is
much more difficult than four states' condition to judge whether the
three states can be distinguished by LOCC. Actually we cannot
provide an analytical characterization. Nevertheless, we can still
get some results qualitatively.

If the equation is satisfied for any $a$ and $b$, we can adjust $a$
and $b$ to make $E_m$ satisfy $\sum E_m=I$. If the equation is
satisfied for a certain value $a_0$ and $b_0$, then we only have an
$E_0=I$. Therefore, Bob can only do a trivial operation on his
system. Then we only need to judge whether these states can be
distinguished if Alice goes first, which is much easier. If there is
no solution to the equation, then the three states cannot be locally
distinguished.

\section{A nontrivial example requiring multi-round classical communication}
Now we turn to discuss the role of classical communication in local
discrimination. We find a set of $m\otimes n$ states needs at least
$2\min\{m,n\}-2$ rounds to be distinguished using LOCC.

First, suppose $m=n$, where $m$ is the dimension of the first
system, Alice's system, and $n$ is the dimension of the second
system, Bob's system. We  construct a set of states as follows:
$$
  \begin{array}{ccccc}
    \ket{0}\ket{\eta_{00}}+&\ket{\alpha_{00}}\ket{0}+&\ket{1}\ket{\eta_{10}}+&\cdots+
    &\ket{n-1}\ket{n-2}\\
    \ket{0}\ket{\eta_{01}},&\ket{\alpha_{01}}\ket{0}, & \ket{1}\ket{\eta_{11}}, & \cdots, & \ket{n-1}\ket{n-1} \\
    \ket{0}\ket{\eta_{02}}, & \ket{\alpha_{02}}\ket{0}, & \ket{1}\ket{\eta_{12}}, & \cdots  &  \\
    \vdots & \vdots & \vdots & \cdots & \\
    \ket{0}\ket{\eta_{0n-1}}, & ~\ket{\alpha_{0n-2}}\ket{0}, & ~\ket{1}\ket{\eta_{10}}, & ~\cdots &~ \\
  \end{array}
$$
where $\{\ket{\eta_{ki}},0\leq i\leq n-k-1\}$ is an orthonormal
basis for the orthogonal complement of
$span\{\ket{0},\cdots,\ket{k-1}\}$ and $\{\ket{\alpha_{li}},0\leq
i\leq n-l-2\}$  is an orthonormal basis for the orthogonal
complement of $span\{\ket{0},\cdots,\ket{l}\}$.
$\ip{\eta_{k1}}{\eta_{l1}}\neq0$,
$\ip{\alpha_{k1}}{\alpha_{l1}}\neq0$, and $\ip{\eta_{k0}}{i}\neq0$
for $k\leq i\leq n-1$, $\ip{\alpha_{l0}}{j}\neq0$ for $l+1\leq j\leq
n-1$. The total number of the states is $n^2-2n+3$.

\begin{theorem}\label{3-special}\upshape The above $n^2-2n+3$ states
need at least $2n-2$ rounds classical communication to be
distinguishable by LOCC.
\end{theorem}

{\bf Proof.} The key idea is to prove that measurement operators
should be projective measurements. Suppose Alice goes first, and let
$E_m$ denote Alice's POVM operator with outcome $m$. As
$\ip{\eta_{01}}{\eta_{k1}}\neq0$ and the orthogonality between
$\ket{0}_A\ket{\eta_{01}}$ and $\ket{k}_A\ket{\eta_{k1}}$ should be
kept after the measurement, we have $\bra{0}E_m\ket{k}=0$,
similarly, $\bra{j}E_m\ket{k}=0$. Therefore $E_m$ is diagonal,
$E_m=diag(\lambda_0,\lambda_1,\cdots,\lambda_{n-1})$.

To keep orthogonality of $\ket{\psi_0}$ and
$\ket{\alpha_{0i}}_A\ket{0}_B$, $E_m$ should also be diagonal under
the bases $\{\ket{\alpha_{0i}}$, $\ket{0}\}$, then
$E_m=\lambda'_0\ket{\alpha_{00}}\bra{\alpha_{00}}+\lambda'_1\ket{\alpha_{01}}\bra{\alpha_{01}}+\cdots+\lambda'_{n-2}\ket{\alpha_{0n-2}}\bra{\alpha_{0n-2}}+\lambda'_{n-1}\ket{0}\bra{0}$.

From the restriction in the theorem, we have
$\ip{\alpha_{00}}{j}\neq0$, for any $j\neq0$. Therefore
$\bra{\alpha_{00}}E_m\ket{j}=\lambda_j\ip{\alpha_{00}}{j}=\lambda'_0\ip{\alpha_{00}}{j}$,
 $\lambda_j=\lambda'_0$, so
$E_m=diag(\lambda_0,\lambda'_0,\cdots,\lambda'_0)$.

If $\lambda_0'$ and $\lambda_0$ are both not 0, then after Alice's
measurement, Bob should do a nontrivial operation on his own system
according to Alice's result. We denote $F_n$ as Bob's operator. As
we discussed above, we can conclude that $F_n$ is diagonal on bases
$\{\ket{0},\ket{1},\cdots,\ket{n-1}\}$. To keep orthogonality of
$\ket{\psi_0}$ and  $\ket{0}_A\ket{\eta_{0j}}_B$, we can also
rewrite
$F_n=\mu'_0\ket{\eta_{00}}\bra{\eta_{00}}+\mu'_1\ket{\eta_{01}}\bra{\eta_{01}}+\cdots+\mu'_{n-1}\ket{\eta_{0n-1}}\bra{\eta_{0n-1}}$.
Following the steps above, as $\ip{\eta_{00}}{j}\neq0$, we have
$\mu_j=\mu'_0$ for arbitrary $j$. Thus $F_n=\mu'_0I$ is a trivial
operator. Finally, either $\lambda_0=0$ or $\lambda'_0=0$.

Notice that this result also suggest that these states cannot be
distinguished if Bob goes first. As we can see the process as Alice
first does a diagonal operator on her system,
$\lambda_0=\lambda_0'=1$. As they are both not $0$, we have proved
that in the above paragraph that after Alice's measurement, these
states cannot be distinguished.

We go back to Alice's first nontrivial measurement. Due to the above
result, Alice's measurement only has two measurement operators:
$E_1=diag(1,0,\cdots,0)$ and $E_2=diag(0,1,\cdots,1)$. If the
measurement outcome is $1$, Bob only needs to do projective
measurements to distinguish the left states. If the measurement
outcome is $2$, the system is then $(n-1)\times n$.

It is then Bob's turn to do measurement. Following the method we
used above , we can similarly prove that Bob's measurement must be
$E_1=diag(1,0,\cdots,0)$ and $E_2=diag(0,1,\cdots,1)$. By induction,
we find the number of rounds needed for distinguishing is $2n-2$.
Hence we complete the proof. \hfill $\square$

In general case, $m\neq n$, we can suppose $m<n$, then to
distinguish the set of states we give in the theorem $2m-2$ rounds
are needed. We can also construct a set of states which requires
$2m-1$ rounds to achieve a perfect discrimination. An explicit
construction is as follows:
$$
  \begin{array}{ccccc}
     \ket{\alpha_{00}}\ket{0}+&\ket{0}\ket{\eta_{00}}+&\ket{\alpha_{10}}\ket{1}+&\cdots
+&\ket{m-1}\ket{m-2}\\
     \ket{\alpha_{01}}\ket{0}, & \ket{0}\ket{\eta_{01}}, & \ket{\alpha_{11}}\ket{1}, &\cdots, & \ket{m-1}\ket{m-1} \\
     \ket{\alpha_{02}}\ket{0}, & \ket{1}\ket{\eta_{02}}, & \ket{\alpha_{12}}\ket{1}, &\cdots, &  \\
    \vdots & \vdots & \vdots & \cdots & \\
    \ket{\alpha_{0m-1}}\ket{0}, & \ket{1}\ket{\eta_{0n-2}}, & \ket{\alpha_{10}}\ket{1}, &\cdots, & \\
  \end{array}
$$
where $\{\ket{\eta_{ki}},0\leq i\leq n-k-2\}$ is an orthonormal
basis for the orthogonal complement of
$span\{\ket{0},\cdots,\ket{k}\}$ and $\{\ket{\alpha_{li}}, 0\leq
i\leq m-l-1\}$ is an orthonormal basis for the orthogonal complement
of $span\{\ket{0},\cdots,\ket{l-1}\}$.
$\ip{\eta_{k1}}{\eta_{l1}}\neq0$,
$\ip{\alpha_{k1}}{\alpha_{l1}}\neq0$, and $\ip{\eta_{k0}}{i}\neq0$
for $k+1\leq i\leq n-1$, $\ip{\alpha_{l0}}{j}\neq0$ for $l\leq j\leq
m-1$.

The proof of the example above is almost the same as the previous
one.

\section{Conclusion}

We have studied the local distinguishablity of $2\otimes 3$ states
when the owner of the qutrit performs the first nontrivial
measurement. We surprisingly find that for certain four or three
states we need to perform the general local POVM in order to achieve
a perfect discrimination, only LPCC is not sufficient. We have
almost completely characterized the local distinguishability of
$2\otimes 3$ states except for some special case when only three
states are under consideration. It would be of great interest to
extend these results to $2\otimes n$ states where $n>3$.

We further construct a special set of $m\otimes n$ states which
require at least $2\min\{m,n\}-2$ rounds classical communication to
finish the discrimination. Our result indicates that classical
communication plays a crucial role in local discrimination. An
interesting open problem is to construct a set of states which may
require more rounds to achieve a perfect discrimination.

We are indebted to the colleagues in the Quantum Computation and
Quantum Information Research Group for many enjoyable conversations.
In particular, we sincerely thank Prof. M. Ying for his numerous
encouragement and constant support on this research. R. Duan is also
grateful to Y. Feng for sharing his insight on LOCC discrimination.
This work was partly supported by the National Natural Science
Foundation of China (Grant Nos. 60621062 and 60503001) and the
Hi-Tech Research and Development Program of China (863 project)
(Grant No. 2006AA01Z102).

\end{document}